\documentclass[twocolumn,preprintnumbers,amsmath,amssymb]{revtex4}

\usepackage{graphicx}
\usepackage{dcolumn}
\usepackage{bm}

\newcommand{\abs}[1]{|#1|}

\usepackage{verbatim}
\begin{document}

\preprint{donorarraytopologicaloptics2019}


\title{Optical and magnetic signatures of the topological edge states in a dimerised donor chain}
\author{Wei Wu}
\email{wei.wu@ucl.ac.uk}
\author{A. J.~Fisher}
\email{andrew.fisher@ucl.ac.uk}
\affiliation{UCL Department of Physics and Astronomy and London Centre for Nanotechnology, University College London, Gower Street, London WC1E 6BT}
\date{\today}%
\begin{abstract}
We have studied the excited states of a one-dimensional donor dimer array by using time-dependent Hartree-Fock and density-functional theories. We find that tuning the inter-donor distances can induce a topological phase transition from a topologically trivial anti-ferromagnetic ground state to a topological phase supporting spin-polarised edge states. Significant changes in the optical spectra accompany this transition, providing signatures for the phase transition and the existence of the edge state; the edge states lead to a robust $1s\rightarrow 2p$ atomic transition. By contrast, the bulk of the chain becomes spinless, inducing a reduction in the HOMO-LUMO gap and a decrease of the excitation energies in the optical spectra. The inclusion of physics beyond the Su-Schrieffer-Heeger-Hubbard model (long-range electron correlations and basis states beyond the 1s manifold) has a critical effect on the  results and leads to the novel prediction of a change in optical response related to a topological phase transition. Our results provide a theoretical foundation for using artificial donor linear arrays in semiconductors to form topological edge states by design. 
\end{abstract}


\maketitle
\textit{Introduction.} The discovery of new states of matter is not only a fascinating branch of fundamental physics, but also plays an important role in the development of new technologies \cite{wen2005}. In the last decade, we have witnessed a rapid development of theory and experiment relating to topological ordering (TO) \cite{hasan2010, qi2011,  wen2017}. A well known example is the fractional quantum hall effect, for which Laudau's spontaneous symmetry-breaking theory is no longer able \cite{wen2017} to interpret the experimental observations, which can be attributed instead to topological ordering. This leads to the new concept of TO without spontaneous symmetry breaking. The one-dimensional spin-1 chain is another important example \cite{haldane1983}, in which the Haldane gap originates from a topological phase. Wen, et. al., \cite{wen2005} have proposed another possible state of matters even beyond TO---quantum ordering \cite{wenbook}. 

The topological insulator (TI) is an important manifestation of TO. The one-dimensional (1D) TI was found almost four decades ago in the linear conjugated polymer, polyacetylene (CH)$_x$, described approximately by the simple Su-Schrieffer-Heeger (SSH) model \cite{sshprl}.  A topological soliton can be found in this model due to the alternating single-double-bonding within the chain of carbon atoms, but electron-phonon interactions are required for this to occur.  The characteristics of the finite 1D TI is the localised charge state at the chain ends, as suggested by a 1D tight-binding model \cite{deplace2011}. In the corresponding infinite periodic structure, the Zak phase \cite{zak1989} is quantized odd (or even) multiples of $\pi$ when the hopping amplitude between unit cells is stronger (or weaker) than that within the cell; the former represents a topologically non-trivial phase while the latter a trivial phase. There have been studies of the inclusion of an on-site Coulomb interaction $U$ into the SSH model, where the main theme is to analyse the effect of the electron correlation on optical properties of the conjugated polymers \cite{soos1993}. However, the phase transition between the topologically trivial and non-trivial states was rarely discussed, perhaps because the topological edge state is not stable in the presence of the electron-phonon interactions within the SSH model.

The original work for two-dimensional TI is the quantum spin Hall effect in 2D \cite{kane2005,bernevig2006}, in which a chiral spin current flows at the edge of a two-dimensional sample without dissipation, while the bulk is insulating \cite{qi2011}. More recently, three-dimensional TIs have been proposed and developed into a stage where they can be classified  \cite{gibney2018,zhang2019,vergniory2018} according to point groups. 

Most attempts to produce TI materials experimentally rely on conventional materials chemistry.  However, in this paper we propose the detection of topological states in arrays of deterministically implanted donors in silicon \cite{schofield2003,stock2020}; these arrays have the advantage that they can be 'frozen' into geometries displaying topological effects by their interactions with the silicon lattice, even if those geometries are not the lowest-energy structures.  Previously we used hydrogen lines as an analogue for lines of donors in silicon within the spherical band approximation \cite{wu2018}. 
Recently, the ground state of uniformly spaced hydrogen chains was studied in detail using advanced strongly-correlated methods including Density Matrix Renormalization Group (DMRG) and Quantum Monte Carlo (QMC) \cite{motta2019}. A variety of phases were found, including a true metal-insulator transition even in one dimension, which could shed important light on the general physics of strongly correlated systems. In addition, finite alternating donor chains were studied theoretically using exact diagonalization within the Hubbard model \cite{le2019}; a crossover was found from the single-particle edge states known in the SSH model at small $U/t$, to a topological magnetic phase at large $U/t$ where the edge states become localised spin states that are weakly bound into a singlet for short chains, and become separate localised spin-1/2 objects in long chains. 

Inspired by Ref.~\cite{motta2019}, here we study a line of ten hydrogen atoms with alternating distances between them to simulate the corresponding donor nanostructures in silicon.  This can be seen as a model going beyond the SSH+$U$ approach \cite{soos1993}, which is obtained by truncating the full many-body Hamiltonian and assuming only short-range interactions. The electron correlations are treated using hybrid-exchange density functionals that have previously shown accurate results for exchange interactions and optical spectra \cite{serri2014, wu2013, zou2018}. In particular, we study the optical excitations of the system across the transition between the trivial anti-ferromagnetic ground state and the topological state using time-dependent density-functional (TDDFT) and Hartree-Fock (TDHF) theories \cite{stratmann1998}.  We find a striking discontinuity of the optical spectra at a certain donor arrangement associated with a collapse in anti-ferromagnetic order and the formation of the spinless bulk state, while spin-polarised edge states emerge simultaneously, leading to a robust and characteristic $1s\rightarrow 2p$ atomic transition. Our calculations show that the competition brought by electronic correlation plays a crucial role for the formation of topological states in realistic one-dimensional systems. 

\begin{figure}[htbp]
\includegraphics[width=9cm, height=0.75cm, trim={1cm 9.5cm 0.0cm 9.0cm},clip]{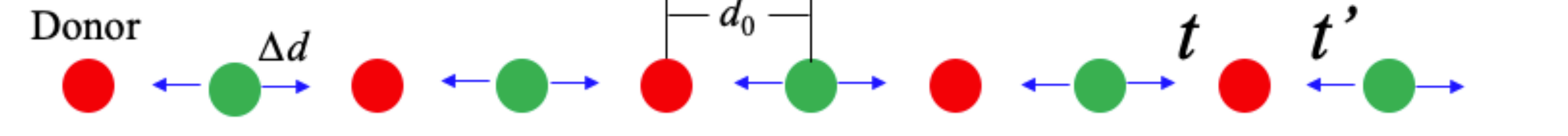}\\
\caption{(Colour online.) The diagram shows the formation of the donor dimer chain studied here. The nearest-neighbouring donor distance for the uniform line is $d_0$. By shifting the donor atoms in green either to left or right by $\Delta d$, we form dimers to probe the topological state. We have also labeled the outer (inner) hopping as $t^\prime$ ($t$).}\label{fig:1}. 
\end{figure}

\textit{Methods}. We used the TDDFT and TDHF methods \cite{stratmann1998} to study the excitations of the system; these were previously shown to be appropriate methods to study the optical properties of donor lines \cite{wu2018}.  TDDFT has been widely used in the computation of excited states in molecules and solids with a great success for interpretation of experimental results. The linear response of charge densities to an external time-domain field, associated with a Dyson-like equation, is used to compute the excitation energies \cite{stratmann1998}. In our TDDFT calculations, we have used a hybrid-exchange functional, in which we tune the proportion of Hartree-Fock exchange in the functional to match the hydrogenic $2s$ and $2p$ excitation energies. The advantage of TDDFT is the better description of the correlation energy as compared with TDHF, although DFT contains a systematic self-interaction error. We have adopted the parameters for donors in silicon, including the effective Bohr radius $a_0^*$ and Hartree (Ha$^*$), within an isotropic single-band effective mass picture as stated in Ref.~\cite{wu2008}. The basis sets include $s$- and $p$-shells as described in Ref.~\cite{wu2018}, where they were used to compute the properties of uniform donor lines. We have computed ten-donor linear arrays while keeping the total spin magnetic quantum number to be zero for all the calculations. The TDDFT modules built in the Gaussian 09 code \cite{g09} were employed to compute the excited states. As shown in Fig.~\ref{fig:1}, we start from a uniform donor line with a donor separation $d_0\sim 12$\,nm when computing the optical spectra; the ground state of this system is expected to be an anti-ferromagnetic (AFM) spin chain. Then we shift alternate donor atoms (blue arrows) by a distance $\Delta d$ (up to $7$\,nm) to form a dimerized linear array. We define $\Delta d > 0$ ($< 0$) when the donors in green (Fig.~\ref{fig:1}) move to the right (left). The broken-symmetry ground state is converged in the self-consistent-field calculations (DFT or HF) as the starting point for the time-dependent calculations. The donor line is arranged along the $z$-direction. To explore the parameter space more fully, we have computed the phase diagram with a set of $d_0$ and $\Delta d$. The quantity we display the phase diagram is $\Delta S = \frac{(\abs{s_1}+\abs{s_{10}})}{2}-\frac{\sum_{i=1}^{8}\abs{s_i}}{8}$, which compares the spin density at the ends of the chain to that in the bulk and can therefore be used to show where the phase transition between AFM and topological states occurs. We show the phase diagram for $8\,\mathrm{nm}<d_0<14\,\mathrm{nm}$ and $0<\Delta d<3\,\mathrm{nm}$, with increments of 0.6\,nm and 0.2\,nm respectively. 

\begin{figure}[htbp]
\includegraphics[width=12cm, height=14.cm, trim={5cm 5.0cm 0.0cm 5.0cm},clip]{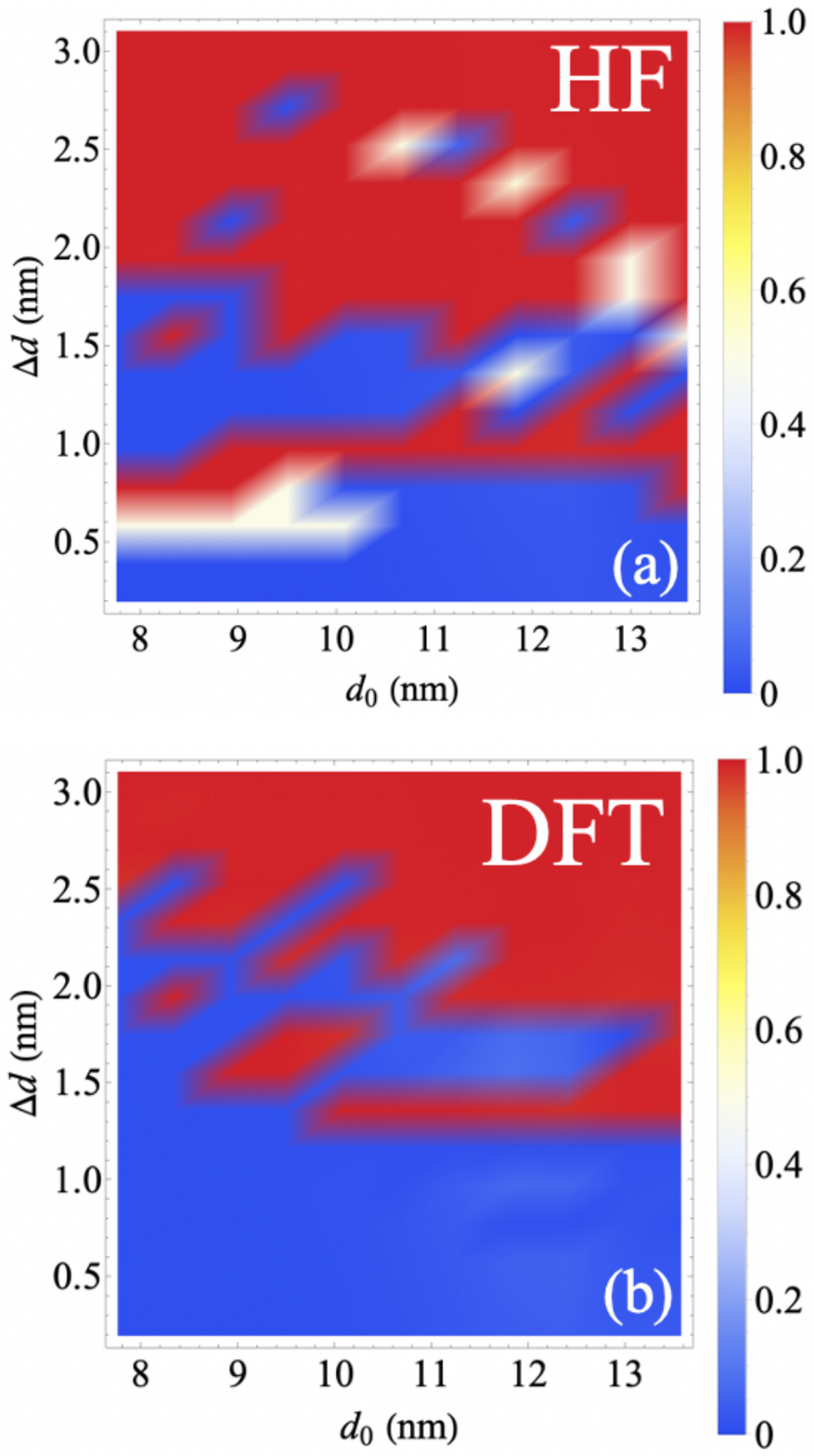}\\
\caption{(Colour online.) Phase diagram of $\Delta S$ as a function of men bond length $d_0$ and shifting distance $\Delta d$ for moving to the right, based on (a) HF and (b) DFT. Red implies a spin-polarised edge state ($\Delta S=1$), whereas blue implies an AFM state ($\Delta S=0$). }\label{fig:2}
\end{figure}

\textit{Hartree-Fock and density-functional-theory ground-state phase diagrams}
We have computed the phase diagram of $\Delta S$ defined in the method section, as a function of the donor distance for the starting uniform chain ($d_0$) and the shifting distance ($\Delta d$). As implied by the definition of $\Delta S$, the system will be in a spin-polarized edge state (AFM state in the sense of mean-field theory) when $\Delta S = 1$ ($\Delta S = 0$). As shown in Fig.~\ref{fig:2}, in a large area of the parameter space we find the topological edge state when $\Delta d$ becomes sufficiently large. From a mean-field point of view (for a Hubbard model), the electron correlation term $U$ is approximately a constant, instead we are effectively varying the ratio between inner and outer transfer integrals ($t$ and $t^\prime$) as a function of $d_0$ and $\Delta d$, as shown in Fig.~\ref{fig:1}. The ratio ($t/t^\prime$) will become larger as we increase $\Delta d$, therefore the spin-polarized edge state, which can be seen as the edge states in a noninteracting tight-binding model split by the on-site Coulomb interaction $U$, will form more easily. We have also performed mean-field-theory calculations (not shown here) for a single-band Hubbard model with alternating transfer integrals, which produce similar spin-polarized edge states within analogous regions of parameter space, consistent with the \textit{ab initio} calculations presented here. We can also see the phase diagram from HF is not uniform at certain parameter area, which could be due to the inter-site electron-electron interactions, or the interactions between $s$ and upper $p$ states, which are absent in a simple single-band Hubbard model. These areas imply the effects of electron-electron interactions on the topological properties in a one-dimensional system.

\begin{figure}[htbp]
\includegraphics[width=12.cm, height=8cm, trim={5.6cm 3.6cm 0.0cm 1.5cm},clip]{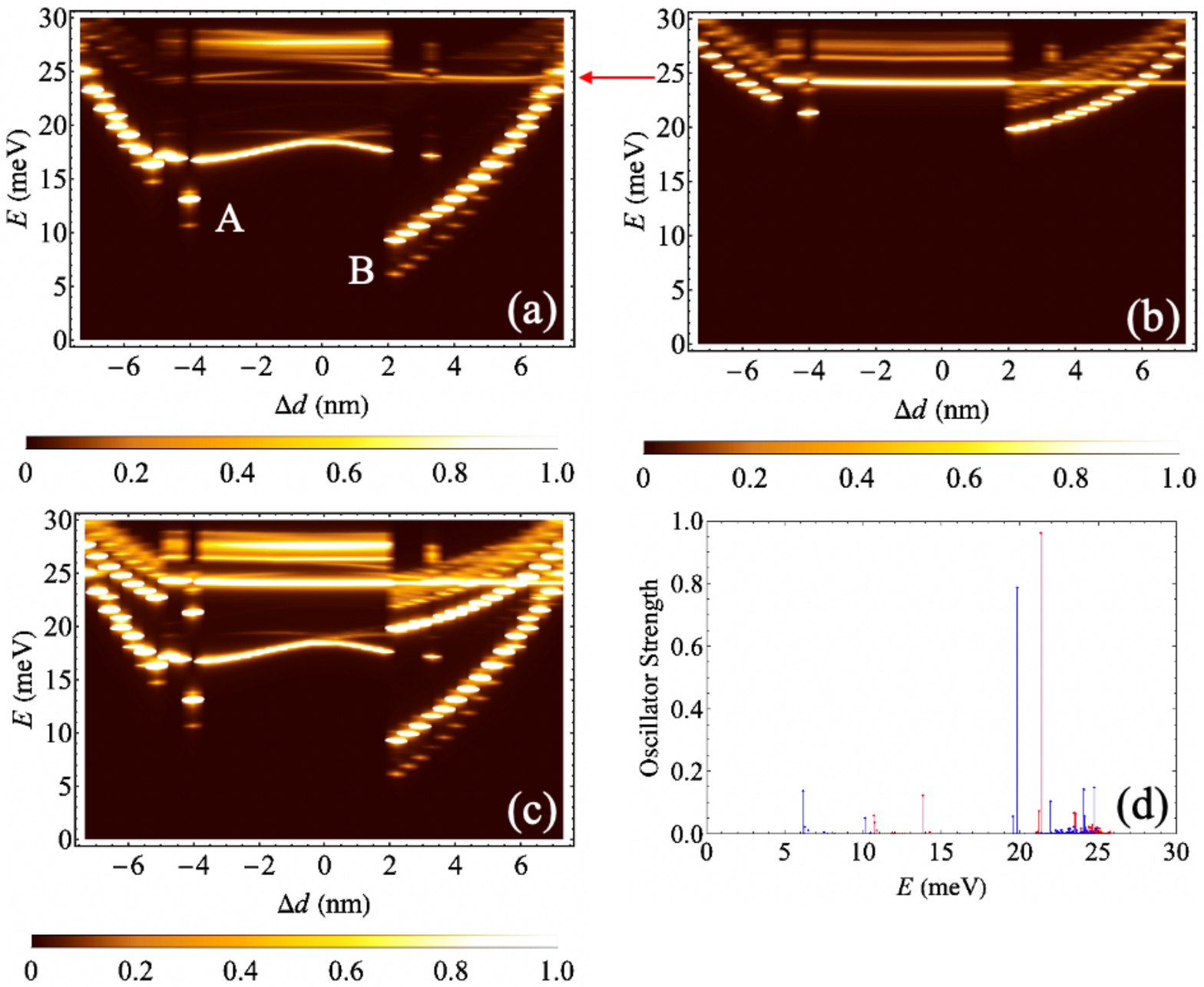}\\
\caption{(Colour online.) Oscillator strengths calculated using TDDFT, as a function of distance shift $\Delta d$ and excitation energy, for a starting uniform donor distance of $d_0\sim 12$ nm: (a) z-component, (b) x- (or y-)component , (c) total oscillator strengths.  In (d), we show the total oscillator strength for positions A ($\Delta d = -4$ nm, red) and B ($\Delta d = 2$ nm, blue), as a function of excitation energy. From (a), we can see the lowest transitions are from the transition along the chain (z-direction), i.e., charge-transfer excitations. The phase transitions from AFM spin chain to a spin-polarized edge state and a non-magnetic dimer state occur at $\Delta d \sim+2$ nm (point B) and $\Delta d \sim-4$ nm (point A) respectively.  The orange arrow in (a) indicates the 1s to 2p transition at 24\,meV.}\label{fig:3} 
\end{figure}

\begin{figure}[htbp]
\includegraphics[width=9.5cm, height=6cm, trim={2cm 2.0cm 0.0cm 2.0cm},clip]{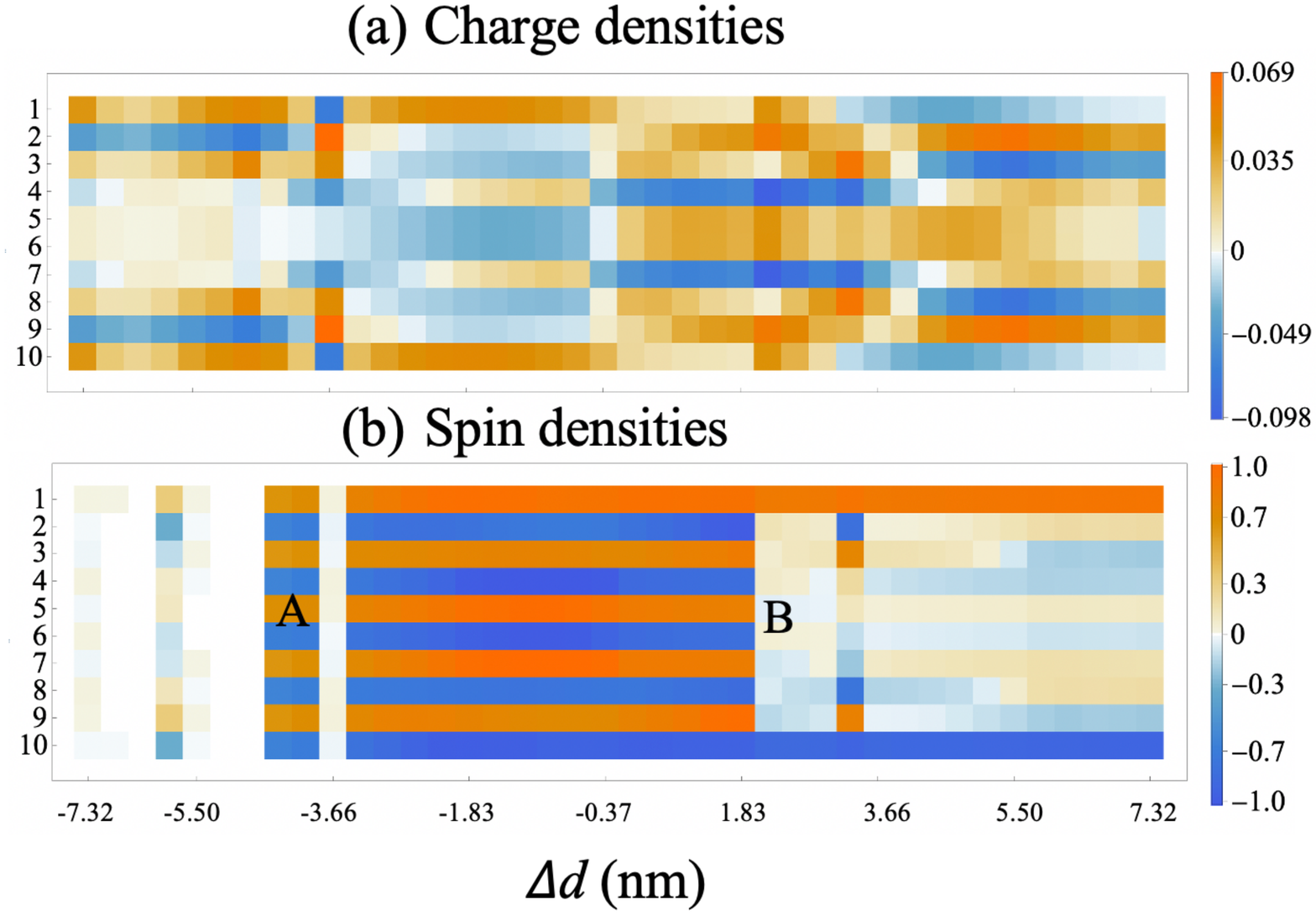}\\
\caption{(Colour online.) (a) Charge densities (red positive and blue negative) and (b) spin densities (red for spin up and blue for spin down) computed within DFT for the configurations in Fig.~\ref{fig:3}. The phase transitions from AFM spin chain to a spin-polarized edge state and a non-magnetic dimer state occur at $\Delta d \sim+2$ nm (point B) and $\Delta d \sim-4$ nm (point A) respectively.}\label{fig:4}
\end{figure}

\textit{TDDFT optical spectra}
As shown in Fig.~\ref{fig:3}, we have computed the oscillator strength based on TDDFT as we shift the donor atoms (in green) to the left and right along the line to form dimers, respectively. We find abrupt changes in the optical spectra at $\Delta d \sim -4$ nm and $\Delta d \sim 2$ nm, labeled by A and B, respectively.  As shown in Fig.~\ref{fig:4}, these are also the points where the AFM order collapses as we move away from $\Delta d$=0.

Fig.~\ref{fig:3}(a) shows the $z$-component (parallel to chain direction) of the oscillator strength; in the central region between the points A and B it is dominated by low-energy excitations varying between 15 and 20\,meV.  We find these are charge-transfer excitations within the bulk antiferromagnetic state; the slight decrease of excitation energies when going away from uniform chains originates from the shorter electron-hole distances in the resulting charge-transfer excitons within each donor dimer. Beyond the transitions at A and B, however, the lowest-energy transitions drop sharply in energy to around 5\,meV at B and 10\,meV at A.  The two lowest branches, both $z$-polarized, are transitions within the bulk of the chain: the lowest transition is between delocalised states, while the second  is predominantly within the dimers.  By contrast, the third lowest branch of excitations is $x$- and $y$-polarized ($\pi$-type excitations), and corresponds to transitions between dimers---see Fig.~\ref{fig:3}(b). 

Although the behaviour in the bulk of the chain, and the corresponding lowest-energy excitations, is qualitatively similar on the A-side ($\Delta d<0$) and the B-side ($\Delta d>0$), the behaviour at the chain ends is very different.  Fig.~\ref{fig:4} shows there is no magnetically polarised edge state for the dimerised chains on the A side ($\Delta d < 0$); instead, we observe some positive charge accumulation on the edge atom of the chain, and negative charge accumulation on the next-to-edge atom.  This is qualitatively similar to the behaviour described in Ref.~\cite{zhu2019}, where carriers are attracted to the deeper potential well in the middle of the chain.  By contrast, on the B side (where $\Delta d<0$, forming dimers in the bulk but leaving a single weakly bonded donor at each chain end) we can see persistent spin-polarised edge states in Fig.~\ref{fig:4}(b) despite the collapse in bulk magnetization; this is accompanied by a robust, persistent optical absorption around 24\,meV (Fig.~\ref{fig:3}), accompanying the appearance of the lower-energy excitations described above.  This 24\,meV absorption corresponds to the single-donor $1s\rightarrow 2p$ transition and is an optical signature of the weakly coupled edge state.

Fig.~\ref{fig:3}(d) shows directly the clear difference between the oscillator strengths as a function of frequency at geometries A and B (just as the bulk magnetization has collapsed in each case).  The clear optical signature of the topological state at B lies in the combination of low-energy transitions below 10\,meV (signalling the collapse of bulk AFM order) with the robust absorption around 24\,meV  (signalling the decoupled edge states.  


\begin{figure}[htbp]
\includegraphics[width=9.5cm, height=5.5cm, trim={1.5cm 2.5cm 0.0cm 2.0cm},clip]{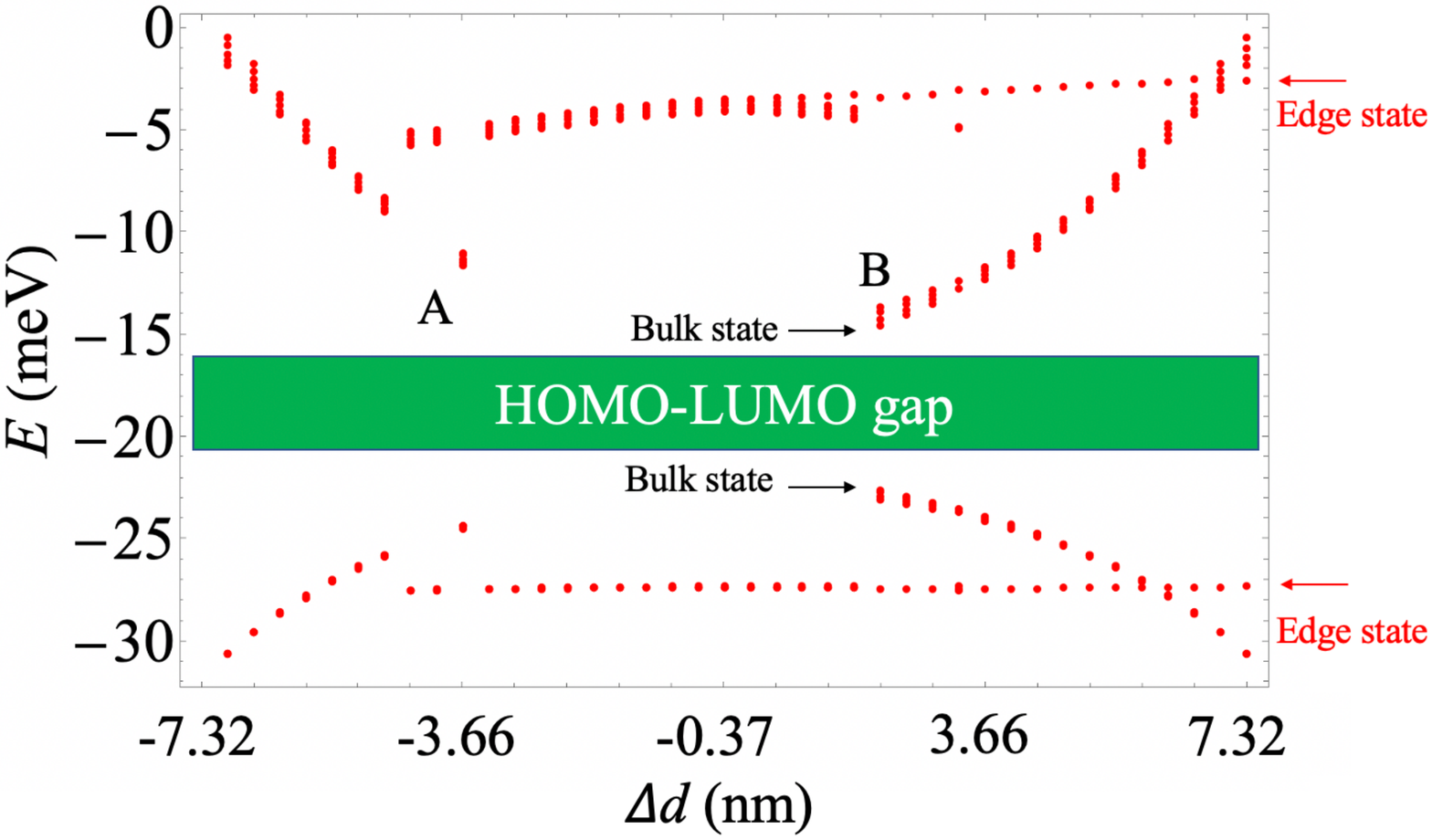}\\
\caption{(Colour online.)  The highest occupied and lowest unoccupied Kohn-Sham energy eigenvalues, computed in hybrid DFT, are shown (the eigenvalues are the same for both spin orientations but the eigenstates are in general different). The phase transition from the AFM spin chain to a spin-polarized edge state happens at $\Delta d \sim +2$\,nm, consistent with the changes in optical spectra calculated in TDDFT. Here A and B label the positions where the phase transitions occur.  The filled and empty edge states are indicated by red arrows.}\label{fig:5}
\end{figure}

We have further analysed the one-electron Kohn-Sham orbital eigenvalues within DFT (Fig.~\ref{fig:5}) in order to gain more insight into the nature of these transitions.  The decrease of the excitation energies at A and B stems from the formation of un-magnetized bulk states which dramatically lower the HOMO-LUMO gap. Although similar abrupt changes in the single-particle energy levels occur at both A and B, edge states can only be seen unambiguously when moving to the right ($\Delta d>0$): a filled edge state is visible around $E=-27$\,meV and an empty one (split by the exchange interaction) at $E=-4$\,meV. 

\textit{Conclusions.} In summary, we have shown that optical excitation could be used as a tool to detect the formation of a topological edge state in a line of hydrogenic donors with alternating bond lengths.  The key signature is the coexistence of a low-lying excitation below 10 meV with a robust atomic-like transition at $\sim 24$ meV.  We have found this signature through performing first-principles TDDFT and TDHF calculations for the ground and excited states of the donor lines. We have also shown that within mean-field theories (HF and DFT) there exists a phase transition between the conventional N\'eel AFM state and a topological state with spin-polarised edge states, as a function of the bond lengths.  

More broadly, our calculations show the effect of electron correlations and long-range Coulomb interactions on the topological properties of a one-dimensional TI originally described by the SSH model. By linking this model explicitly to a physically realisable system (dopants in semiconductors) and a measurable quantity (the optical absorption), this work would provide a theoretical foundation to use semiconductor donor arrays to probe the topological properties of quantum matter and demonstrate an artificial 'designer' topological insulator.

It is important to consider potential qualifications to our results.  First, the Mermin-Wagner theorem \cite{mermin} leads us to expect there should be no long-range order in two dimensions or below at finite temperatures in a spin model with a broken continuous symmetry and short-range interactions, owing to long-wavelength thermal fluctuations.  Even at zero temperature we would expect such order to be suppressed by quantum fluctuations in one dimension. By contrast, our calculations show static AFM order throughout the chain near $\Delta d=0$, and a fixed relationship between spins at the chain ends for $\Delta d\ge+2\,\mathrm{nm}$; it is likely that the true states involve a superposition of different orderings, giving rise to magnetic correlations but no static ordering, as found in the corresponding Hubbard model with short-range interactions \cite{le2019}. Second, the predictions for the optical signatures presented here are within an isotropic single-valley approximation (the hydrogenic case); for donor pairs, we have recently shown \cite{wu2020} that including the conduction-band anisotropy and multi-valley effects does change the optical absorption, but preserves the essential features of (i) charge-transfer transitions at small distances, and (ii) atomic-like 1s to 2p transitions at large distances.  We therefore expect that the main features of our proposed optical signature will survive in a multi-valley treatment, even in the presence of a superposition of possible magnetic orderings.

\begin{acknowledgments}
We wish to acknowledge the support of the UK Research Councils under Programme Grant EP/M009564/1. We thank Nguyen H. Le, Eran Ginossar, Gabriel Aeppli, Neil Curson, Taylor Stock, Alex K\"{o}lker, Guy Matmon and Cedric Weber for helpful and inspiring discussions.
\end{acknowledgments}


\end{document}